\documentclass[preprint]{vgtc}                 %

\ifpdf%
  \pdfoutput=1\relax                   %
  \usepackage{graphicx}                %
  \DeclareGraphicsExtensions{.pdf,.png,.jpg,.jpeg} %
\else%
  \usepackage{graphicx}                %
  \DeclareGraphicsExtensions{.eps}     %
\fi%

\graphicspath{{figures/}{pictures/}{images/}{./}} %

\usepackage{microtype}                 %
\PassOptionsToPackage{warn}{textcomp}  %
\usepackage{textcomp}                  %
\usepackage{mathptmx}                  %
\usepackage{times}                     %
\usepackage{cite}                      %
\usepackage{tabu}                      %
\usepackage{booktabs}                  %

\onlineid{2343}

\vgtccategory{Application}

\vgtcinsertpkg

\preprinttext{2021 IEEE Virtual Reality and 3D User Interfaces (VR)}
\usepackage{enumitem}
\newcommand{\compresslist}{
  \setlength{\itemsep}{1pt}
  \setlength{\parskip}{0pt}
  \setlength{\parsep}{0pt}
}
\usepackage{tabularx}
\usepackage{mathtools}
\usepackage{balance}

\newcommand{\comm}[1]{}

\title{Towards Sneaking as a Playful Input Modality for Virtual Environments}

\author{Sebastian Cmentowski\thanks{e-mail: sebastian.cmentowski@uni-due.de}\\ %
        \parbox{1.8in}{\scriptsize \centering High Performance Computing Group \\ University of Duisburg-Essen} %
\and Andrey Krekhov\thanks{e-mail: andrey.krekhov@uni-due.de}\\ %
     \parbox{1.8in}{\scriptsize \centering High Performance Computing Group \\ University of Duisburg-Essen\\~\\} %
\and André Zenner\thanks{e-mail: andre.zenner@dfki.de}\\ %
     \parbox{1.8in}{\scriptsize \centering Saarland University \& DFKI \\
Saarland Informatics Campus} %
\and Daniel Kucharski\\ %
     \parbox{1.8in}{\scriptsize \centering High Performance Computing Group \\ University of Duisburg-Essen} %
\and Jens Kr\"uger \thanks{e-mail: jens.krueger@uni-due.de}\\ %
     \parbox{1.8in}{\scriptsize \centering High Performance Computing Group \\ University of Duisburg-Essen}}

\abstract{%
Using virtual reality setups, users can fade out of their surroundings and dive fully into a thrilling and appealing virtual environment. The success of such immersive experiences depends heavily on natural and engaging interactions with the virtual world. As developers tend to focus on intuitive hand controls, other aspects of the broad range of full-body capabilities are easily left vacant. One repeatedly overlooked input modality is the user's gait. Even though users may walk physically to explore the environment, it usually does not matter how they move. However, gait-based interactions, using the variety of information contained in human gait, could offer interesting benefits for immersive experiences. For instance, stealth VR-games could profit from this additional range of interaction fidelity in the form of a sneaking-based input modality.

In our work, we explore the potential of sneaking as a playful input modality for virtual environments. Therefore, we discuss possible sneaking-based gameplay mechanisms and develop three technical approaches, including precise foot-tracking and two abstraction levels. Our evaluation reveals the potential of sneaking-based interactions in IVEs, offering unique challenges and thrilling gameplay. For these interactions, precise tracking of individual footsteps is unnecessary, as a more abstract approach focusing on the players' intention offers the same experience while providing better comprehensible feedback. Based on these findings, we discuss the broader potential and individual strengths of our gait-centered interactions.
} %

\CCScatlist{
  \CCScatTwelve{Human-centered computing}{Virtual reality}{}{};
  \CCScatTwelve{Software and its engineering}{Interactive games}{}{}
}

\teaser{
  \centering
  \includegraphics[width=\linewidth]{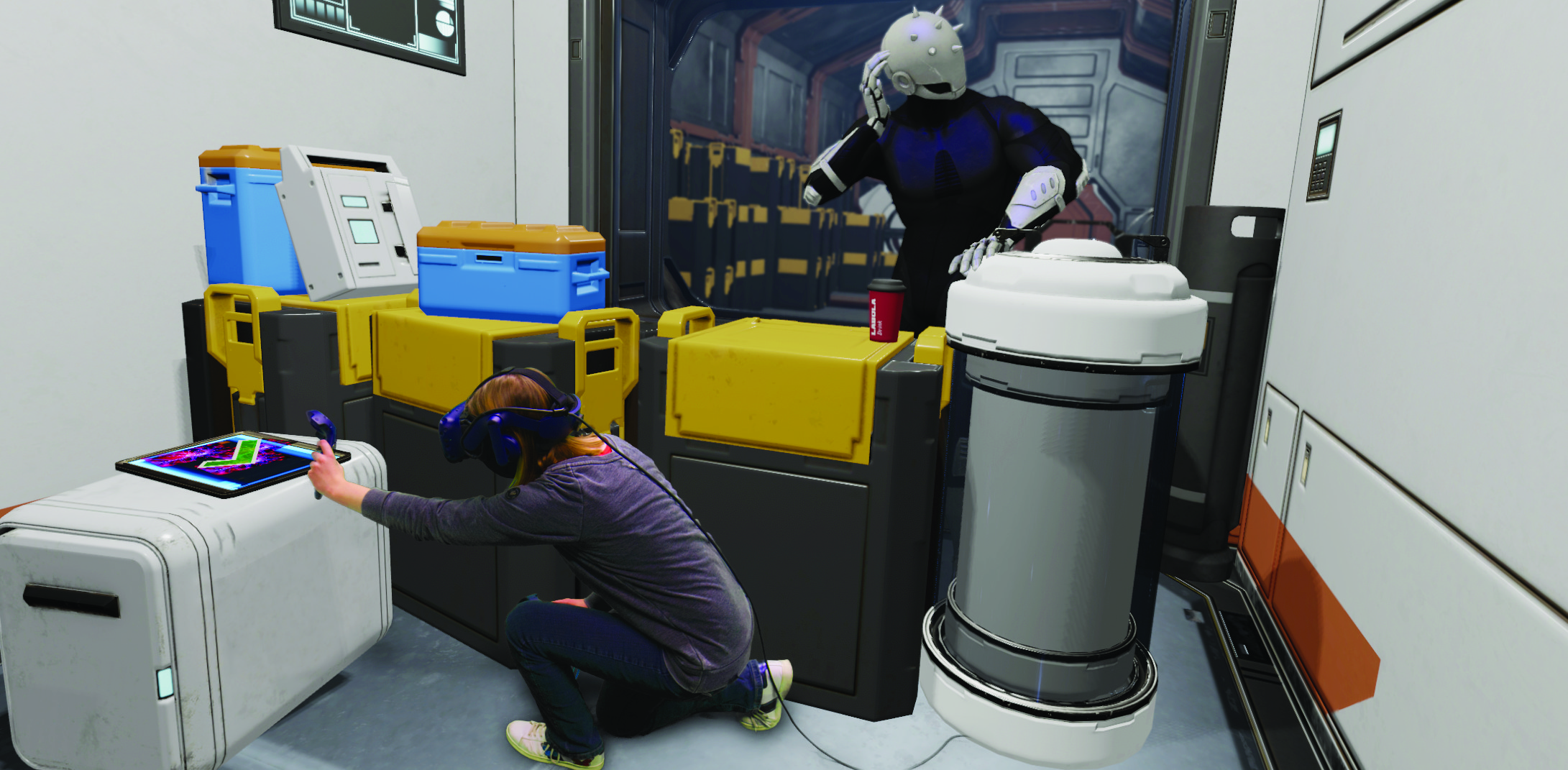}
  \caption{We explore the potential of sneaking as a novel input modality for immersive virtual environments. In addition to hiding visually, players also must pay attention to their own gait to avoid detection.}
	\label{fig:teaser}
}

\begin{document}

\firstsection{Introduction}

\maketitle

Imagine being a spy infiltrating a secret base, sneaking past patrolling guards, stealing the confidential information, and leaving --- unseen. This plot reads like a typical mission of any stealth game. While such games already deliver an intense experience when consumed on a flat-screen, virtual reality (VR) setups provide the unique potential of boosting tension and involvement even further. Players may fully dive into the character's role and experience the plot themselves. Nevertheless, existing stealth VR-games often fail to reach this enormous potential. Most of the available titles, such as Espire 1: VR Operative~\cite{GameEspire}, do not offer a fully fetched sneaking mechanism. Instead, players have to use virtual locomotion techniques and activate a binary sneak mode using hardware buttons. While enemies may visually detect the players, one of the central aspects of sneaking is left vacant: being quiet. 

In the real world, every step we take emits noise. Apart from revealing our position and speed, these walking sounds also expose a broad range of personal information, including gender~\cite{li1991perception}, emotional state~\cite{giordano2006walking}, or posture~\cite{pastore2008auditory}. Still, we usually tend to ignore them in our everyday life. Things change when the situation requires secrecy and discretion. Sneaking needs both -- staying out of sight and adopting the own gait to minimize walking sounds. On the other hand, one can also attract attention by producing sounds intentionally, i.e., through stomping. This range of interaction fidelity, which is still missing in today's VR games, could greatly benefit immersive VR experiences.

Our research closes this gap by assessing the potential of sneaking as a novel input modality for immersive virtual environments (IVEs). Therefore, we split our work into three consecutive parts: We start by developing the technical basis for capturing the users' sneaking behavior. In this process, we provide insights into our exploratory design process that covered a range of different technical approaches, e.g., using microphones or marker-based tracking, as well as various abstraction levels. We also discuss the reasoning behind our final selection of three fundamentally different implementations.

In the second part of this work, we develop possible interactions and gameplay elements utilizing our stealth mechanisms. Combining sneaking with other time- or body-based tasks allows us to modify the overall task difficulty and provide varied challenges. In the final step, we compare our three sneaking mechanisms in a between-subject study, using the developed interaction concepts. Our context is an immersive stealth VR-game, where the subjects take the role of a secret agent, stealing confidential information (see Figure~\ref{fig:teaser}). 

Our results reveal the great potential of sneaking as an input modality for IVEs, providing high presence and enjoyment levels. The body-based mechanisms offer unique challenges and thrilling gameplay. Comparing the different technical approaches, we found that precise tracking of individual footsteps is unnecessary for the particular use-case. Instead, a more abstract approach focusing on the players' intention offers the same experience while providing better comprehensible feedback. In turn, accurate footstep tracking could be used for a variety of other use-cases, e.g., for training simulations providing individual gait-related feedback. These findings form the basis for further research on other gait-related interaction concepts.

\section{Related Work}

In this section, we cover the related research relevant to this work. We start by briefly covering the basic concepts linked to playing games in VR --- immersion, presence, and cybersickness. Next, we provide a concise summary of VR locomotion research, as our topic is closely linked to this area. Lastly, we discuss the latest advancements enriching IVEs with novel sensations or a greater input fidelity, focusing primarily on gait-related approaches.

Modern head-mounted displays (HMDs), such as the Oculus Quest 2~\cite{oculus}, replace the users' real surroundings with a realistic representation of the virtual world. In this context, researchers usually call the technical quality of the used hardware \textit{immersion}~\cite{cairns2014immersion, Biocca:1995:IVR:207922.207926, sherman2002understanding} and the perceptual effect of being in the IVE \textit{presence}~\cite{ heeter1992being, slater2003note}. The latter is of particular interest for this work and can be measured using various approaches~\cite{UQO.2004, IJsselsteijn, lombard1997heart}. Apart from the positive experience of diving into a fully immersive environment, VR also bears the risk of causing cybersickness~\cite{laviola2000discussion, hettinger1992visually}. In this case, a mismatch between the human vision and the vestibular system causes symptoms ranging from headaches to vomiting.

Especially poorly suited locomotion techniques bear the risk of quickly inducing high levels of discomfort~\cite{ Habgood:2017:HLP:3130859.3131437}. Thus, recent work has mostly focused on using natural locomotion approaches, such as real walking~\cite{ruddle2009benefits}, to counter this threat. These efforts center around the concept of extending the walking range by augmenting real movements~\cite{bolte2011jumper, Bhandari:2017:LSW:3139131.3139133, interrante2007seven}, changing perspectives~\cite{GulliVR, cmentowski2019outstanding, abtahi2019m}, or subconsciously avoiding real obstacles~\cite{razzaque2001redirected}.  Among these approaches, the \textit{walking-in-place} concept is of particular interest as it derives the users' anticipated motion from in-place steps~\cite{slater1995virtual}. While being superior to gamepad locomotion, early implementations did not feel as natural as real walking~\cite{usoh1999walking}. Thus, later research~\cite{feasel2008llcm, templeman1999virtual, yan2004new} has focused on improving the matching, e.g., by using the biomechanics of human gait~\cite{wendt2010gud}. Since our work connects only loosely to VR locomotion research, we point to Boletsis et al.~\cite{boletsis2017new} and Krekhov et al.~\cite{krekhov2020player} for a more detailed overview of the current state of the art.

Recently, a particular focus in VR research has been placed on enhancing the users' perception of the virtual world. These efforts mostly focus on adding additional sensations exceeding the visual and audio components of currently available headsets. Examples include haptic surface feedback~\cite{whitmire2018haptic}, weight-shifting controllers~\cite{zenner2019drag}, or olfactory systems~\cite{nakamoto2020virtual}. Instead of adding novel sensations, other projects used the existing capabilities to manipulate the users' impression of the virtual world. In this context, a special focus was placed on the effects of displaying virtual avatars in various shapes and appearances~\cite{lugrin2018any}. For instance, altering the users' avatar can not only evoke the impression of changing age~\cite{banakou2013illusory}, race~\cite{kilteni2013drumming, peck2013putting}, or even species~\cite{krekhov2019beyond} but also impact mental health, like in the case of eating disorders~\cite{perpina1999body}. Specifically related to our particular research interest, Pan and Steed~\cite{pan2019foot} focused on the influence of virtual legs and feet on presence and embodiment.

Apart from enriching the users' interactions and sensations in general, a growing body of research has focused on improving the movement through virtual scenarios. A compelling and realistic walking experience requires a profound knowledge of the biomechanical fundamentals of human gait. The upright bipedal progression characterizing human locomotion is commonly defined as a periodic movement of the two lower limbs. The underlying pattern, i.e., the \textit{gait cycle}~\cite{inman1981human}, consists of two phases: the swing phase, where the foot is in the air, and the stance phase, where the foot has contact with the ground~\cite{kharb2011review}. The latter phase is subdivided into four stages~\cite{huang2006intelligent, morley2001shoe}: heel strike, forefoot contact, midstance, and heel off. The interplay of alternating swing and stance phases leads to forward propulsion. Individual differences, such as gender, age, posture, or walking speed, significantly influence the particular gait~\cite{ko2010characteristic, troje2008retrieving}, and it was shown that listeners could deduce these characteristics solely from the emitted walking sounds~\cite{visell2009sound}. Also, research mainly differentiates between two primary types of gait: walking and running~\cite{diedrich1995change, hreljac2007does}. Apart from the velocity, the main difference is the flight phase while running, i.e., no foot is touching the ground. Thus, we consider sneaking a subtype of walking, characterized by a more careful foot placement.

In the last decades, foot-based interactions have been of ongoing interest to the research community~\cite{velloso2015feet}. For virtual scenarios, most of the work has focused on more realistic walking experiences. The \textit{Real-Walk} approach by Son et al.~\cite{son2018realwalk} simulates various surfaces by altering the viscosity of a shoe-like apparatus. Similarly, King et al.~\cite{terziman2012king} combined visual with tactile vibrations to improve the realism of walking in VR. Strohmeier et al.~\cite{strohmeier2020barefoot} presented their \textit{bARefoot} prototype capable of generating virtual walking surfaces through motion-coupled vibrations. Our work --- exploring sneaking as an input modality --- adds to the research dealing with the auditive aspect of walking. This research area is mostly centered around synthesizing or modifying the users' footstep sounds as part of the overall soundscape. For instance, Tajadura et al.~\cite{tajadura2015light} showed that modified walking sounds alter the self-perception. Also, Kern et al.~\cite{kern2020audio} reported the positive effects of synchronized step sounds on presence and realism. For a broader view on the interplay between synthesized footstep sounds and immersive soundscapes, we point to the extensive work by the Medialogy Department at Aalborg University Copenhagen~\cite{serafin2009extraction, nordahl2011sound, turchet2010examining}. The closest related work in this field is the \textit{VRSneaky} approach by Hoppe et al.~\cite{hoppe2019vrsneaky}. The authors use shoe-attached trackers to play gait-aware walking sounds in a stealthy IVE to provoke a gait-change and achieve an increased presence. While these works underline the importance of plausible soundscapes, including synchronized walking sounds, our work concentrates on the potential of the users' sneaking behavior as a novel input modality.

\section{Developing the Sneaking Mechanisms}

While past research has already dealt with capturing the users' walking behavior and using this information in virtual environments, these approaches mainly aimed to achieve a realistic soundscape. In contrast, our mechanism should translate the users' gait into a discrete state, i.e., differentiating between walking and sneaking. This processed information would form the basis for our novel input modality. From the very start of our design process, we decided to start from scratch and refrain from using any auditory walking feedback with our mechanism. Existing research, such as VRSneaky~\cite{hoppe2019vrsneaky}, have already demonstrated the benefits of precisely synchronized footstep sounds for presence and gait awareness. We wanted to focus entirely on the interactional aspect and determine the necessary tracking fidelity needed for a plausible sneaking mechanism.

Based on this goal, we determined three main requirements for the target implementation. Firstly, the mechanism must be able to differentiate between the two states \textit{walking} and \textit{sneaking}. We do not distinguish between walking and stomping, as both share the same source, i.e., stepping with normal force, and the same effect, i.e., attracting attention. Secondly, the used tracking mechanism must deliver robust signals to determine the active state independent of the users' physiologies, walking behavior, and other environmental factors, i.e., ground, footwear, or noise. Finally, the chosen tracking method must not impede the users' movement in the real world in any way. This constraint also applies to hardware that might alter or diminish the fine-graded foot movement necessary for sneaking. 

Considering these prerequisites, we started by capturing the real step sounds emitted by the users. Therefore, we attached Bluetooth microphones to the users' ankles and used the transmitted audio volume to extract the users' gait. This approach is the exact realization of the abstract idea behind our work. The lightweight microphones guarantee an easy setup that is not intervening with the actual gameplay. However, external interferences and individual differences between users are only partially removable by filtering. Eager to find a better alternative, we experimented with various other approaches. In particular, we shifted our focus from measuring the exact sounds to detecting the foot motions during sneaking.

Compared to other alternatives, such as force-sensing resistors, our final implementation uses the existent precise VR tracking environment and only requires a pair of HTC Vive trackers attached to the users' feet. This setup does not influence the individual sneaking movement and provides seamless and quick integration with the overall VR system. As with every sensor device, minor tracking errors and inaccuracies might occur from time to time. Also, using the trackers' exact positions to determine the foot's touchdown would require precise calibration as every users' feet are different. However, we found that these issues are avoidable by rethinking the definition of silent footsteps. When a foot is placed on the floor, the ground slows its speed to zero. The faster a foot is slammed down, the more noise is produced. Thus, the step sounds are directly dependent on the decrease in velocity.

\begin{figure}[tb]
 \centering %
 \includegraphics[width=\columnwidth]{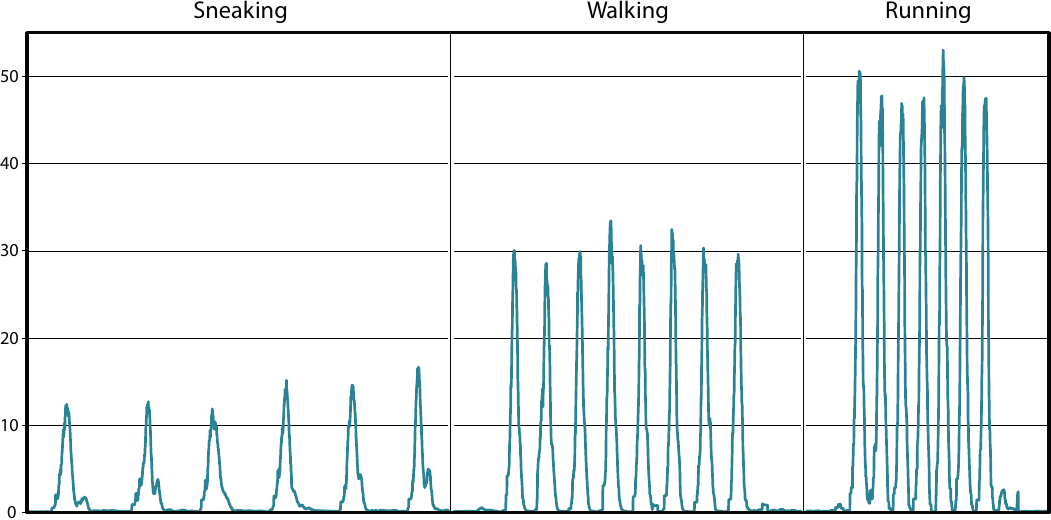}
 \caption{Plotted deceleration values (in $m/s^2$) for a typical player. The average peaks for sneaking, walking, and running activities (from left to right) vary only minimally across different users, providing robust thresholds for the different types of gait.}
 \label{fig:detection}
\end{figure}

Plotting the decrease in velocity over time reveals the users' gait. Note that we isolate the deceleration alone:  
\[
    d(v,t)= 
\begin{dcases}
    \frac{\Delta\, v}{\Delta\, t}& \text{if }\, \Delta\, v \leq 0\\
    0              & \text{otherwise}
\end{dcases}
\]
The resulting peaks correspond to when the users place their foot on the floor. These measured values vary only minimally across different users, making it possible to determine global thresholds for different types of gait (see Figure~\ref{fig:detection}). While this approach is immune to typical noise sources, such as random foot movements or physiological differences, we noticed that minor tracking errors might still trigger an unexpected peak. We solved this impediment by adding a small sliding window to eliminate single erroneous frames. After all, this tracker-based approach has a minimal impact on the movement and provides a reliable and precise tracking of the users' gait.

During the design process of our first implementation, we learned that it might be beneficial to use a more abstract approach instead of measuring walking sounds directly. The result stays the same: we can detect whether users are walking or sneaking. Next, we asked ourselves: Do we even need to track individual steps? When testing our early prototypes, all test users shared one similarity when sneaking: they walked slowly. While it would certainly be possible to sneak quietly without sacrificing much of the original walking speed, we usually measured a significant decrease in general velocity during our trials. It seems that this observation is tightly associated with the users' expectations. Thus, we designed our second approach to measure the overall walking speed using the horizontal velocity of the HMD. The threshold, determining whether users sneak or walk, was chosen based on the tracking data of our pre-tests. Like the first tracker-based mechanism, we added a sliding window to account for tracking issues and the head's micro-movements. Apart from that, both approaches work almost identical.

\begin{figure*}[tb]
 \centering %
 \includegraphics[width=\textwidth]{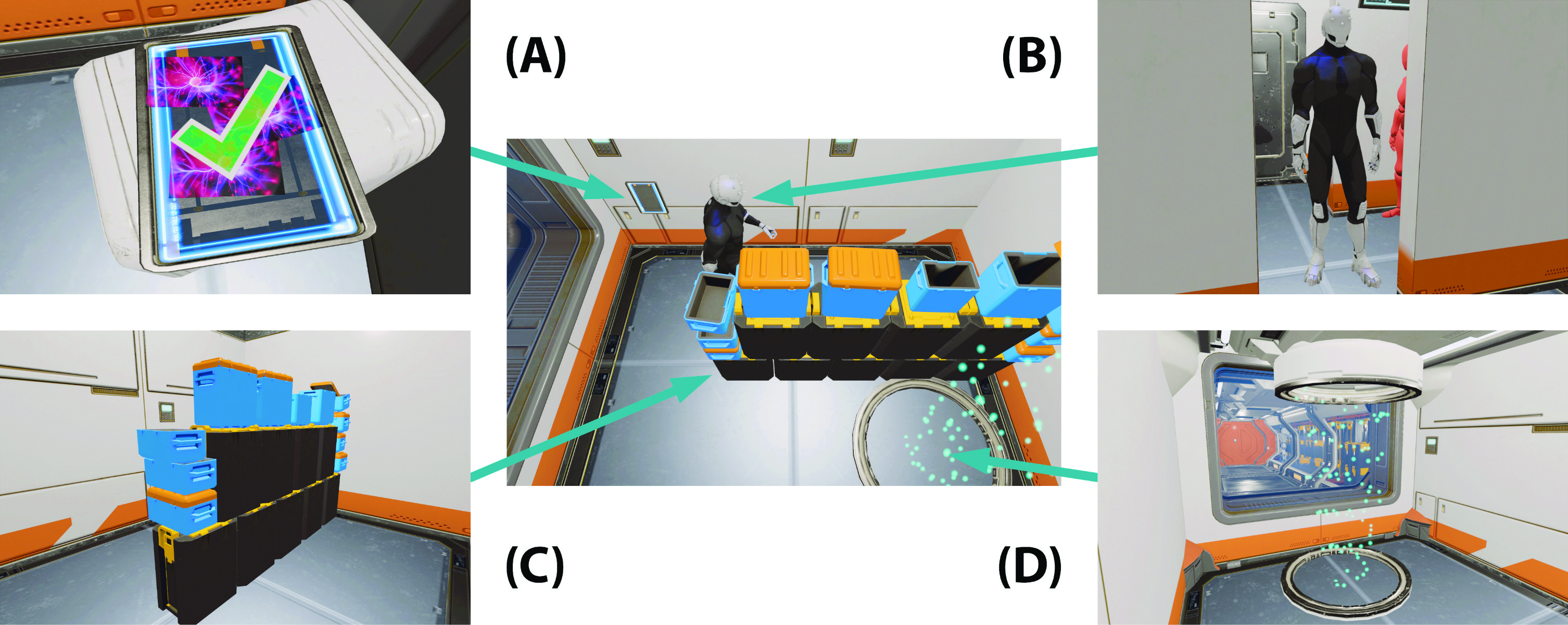}
 \caption{Overview of a typical level of our testbed game, including all important elements: the target tablet (A), the enemy guard (B), obstacles providing cover (C), and the teleporter serving as entry and exit (D).}
 \label{fig:scenario}
\end{figure*}

The main difference between our two implementations is the loss of tracking fidelity. Only the first technique can detect the actual gait, whereas the second mechanism relies on an implicit observation. Nonetheless, both approaches are body-based interactions and are compatible with natural walking, which is generally seen as the best locomotion technique for virtual environments~\cite{usoh1999walking}. In contrast, existing stealth VR-games are geared to sneaking mechanisms of established non-VR games. These games tend to use a binary stealth mode that is triggered using a hardware button. Players are slowed down and become harder to detect by the NPCs. Of course, this approach requires a virtual locomotion technique to control the player's position and movement. While not entirely comparable to our walking-oriented mechanisms, we decided to add gamepad-based sneaking as a third baseline implementation. This approach uses a continuous joystick movement and introduces a dedicated button triggering the sneaking mode. While sneaking, users are limited to a lower velocity that does not attract attention. It is worth noting that this mode is optional, as users might use the joystick carefully enough to achieve the same effect.

Altogether, our design and implementation phase leaves us with three different approaches to detect sneaking:
\begin{itemize}[leftmargin=*]\compresslist
\item \textit{Tracker}: ankle-attached trackers measure the foot's deceleration
\item \textit{HMD}: HMD's movement is used as a proxy for the users' speed
\item \textit{Gamepad}: joystick locomotion and button-controlled sneak mode
\end{itemize}

\section{Designing Sneaking-based Interactions}

Apart from developing the technical basis for measuring the users' gait, it is crucial to design gameplay elements and interactions that utilize the novel input channel. On the surface, our sneaking mechanism appears to be a simple binary differentiation: If the players walk carefully enough, everything is fine. Otherwise, they are detected.

In this regard, sneaking is very similar to hiding visually from enemies. Both mechanisms build on top of the normal locomotion in the virtual scenario. However, they differ in the underlying challenge. Hiding behind obstacles and avoiding a direct line of sight is a complex body-based task. Players must pay attention to the positioning of all body parts, i.e., not letting a limb peek out of the coverage. On the other hand, sneaking does not constrain the players' position but requires careful movement. Thus, hiding is challenging the \textit{where to move} and sneaking the \textit{how to move}. Both mechanisms are often used in conjunction to increase the overall difficulty.

Besides this famous coalescence, sneaking might be combined with other interaction patterns to obtain interesting challenges and vary the degree of difficulty. We subdivide these patterns into two groups: \textit{reinforcing} and \textit{contradicting} elements. A task is called \textit{reinforcing} if it intensifies the players' attention on their movement. Typical examples include stepping over obstacles or crouching behind barriers. These tasks add an additional movement challenge forcing the players to focus even more on every step. Conversely, \textit{contradicting} gameplay elements split the players' attention between the newly emerged challenge and the active sneaking task. For instance, getting past patrolling guards requires spatial understanding and precise timing, while players still must avoid getting heard. 

\setlength{\tabcolsep}{2pt}
\begin{table}[t]
  \caption{Gameplay overview of our testbed stealth game. In each of the ten consecutive levels, we add one additional gameplay concept.}
  \label{tab:levels}
  \begin{tabularx}{\columnwidth}
  {>{\raggedright\arraybackslash}p{0.7cm}
  >{\raggedright\arraybackslash}X
  >{\raggedright\arraybackslash}p{3.2cm}
  }
    \toprule
    Level & Description & Gameplay Concept \\
    \midrule
    01 & guard is outside of the room & steal the tablet\\
    02 & guard is outside, tablet is hidden & search the tablet\\
    03 & guard is facing the wall & sneak\\
    04 & guard is observing the obvious path  & avoid the guard's sight\\
    05 & guard is turning regularly & time the own movements\\
    06 & guard patrols along a fixed path & analyze the patrol\\
    07 & guard patrols through the entire room & keep moving\\
    08 & dynamic obstacles provide cover & combine multiple timings\\
    09 & lasers block the path at breast height  & crouch past obstacles\\
    10 & guard stands behind a counter & move while crouching\\
    \bottomrule
\end{tabularx}
\end{table}

Finally, the switch between quiet sneaking and loud walking itself is another potential gameplay element. A common feature found in almost every stealth game is the capability of producing sound on purpose. Whistling or stomping may attract attention and lure enemies to desired spots. In essence, this interaction combines the known gameplay elements. Stomping is another type of \textit{reinforcing} action, as it requires stopping sneaking for one single step. Next follows a typical \textit{contradicting} time-based task, e.g., guards investigating the noise's source. Thus, we consider such composite features to be more complex than their primary counterparts.

\section{Evaluation}

We conducted a study to evaluate our three proposed sneaking techniques using a between-subject design. We were primarily interested in the general acceptance of our mechanisms and the differences in enjoyment, presence, tension, difficulty, and necessary effort. Therefore, we designed a VR stealth game that consisted of multiple short levels, each focusing on one of the introduced interactions. 

\subsection{Research Questions and Hypotheses}
The study's main goal is to explore the differences between the three implementations Tracker, HMD, and Gamepad. In this context, we refer to both of our two proposed approaches, HMD and Tracker, as full-body movement-based interactions. We hypothesize that these techniques, resembling real sneaking, benefit the perceived presence. Additionally, such gait-based interactions force the players to pay attention to every step they take. Consequently, we expect a significantly higher physical effort compared to the Gamepad controls. Also, we assume an overall increase in task complexity. Despite these difficulties, we do not expect a lower success rate. Instead, the novel challenges are likely to increase the players' tension and enjoyment. All in all, our hypotheses are as following:

\setlength{\tabcolsep}{2pt}
\begin{table*}[t]
  \caption{Mean scores, standard deviations, and one-way ANOVA values of the Presence Questionnaire (PQ), the Intrinsic Motivation Inventory (IMI), the Player Experience Inventory (PXI), the NASA Task Load Index (NASA-TLX), and the Simulator Sickness Questionnaire (SSQ).}
  \label{tab:quest}
  \begin{tabularx}{\textwidth}
  {>{\raggedright\arraybackslash}p{4cm}
  >{\centering\arraybackslash}X
  >{\centering\arraybackslash}X
  >{\centering\arraybackslash}X
  >{\centering\arraybackslash}p{1.5cm}
  >{\centering\arraybackslash}p{1.5cm}
  >{\centering\arraybackslash}p{0.6cm}
  >{\raggedright\arraybackslash}p{0.3cm}
  }
    \toprule
     & Tracker ($N = 15$) & HMD ($N = 15$) & Gamepad ($N = 15$) & F(2,42) & $\eta^{2}$ & $p$ & \\
    \midrule
    PQ (scale: 0 - 6)\\
    \ \ \ \ \ \ Realism & 4.98 (0.63) & 5.03 (0.68) & 4.02 (0.91) & 8.695 & 0.293 & .001 & **\\
    \ \ \ \ \ \ Possibility to Act & 4.92 (0.71) & 4.78 (0.55) & 4.83 (0.69) & 0.161 & 0.008 & .852 &\\
    \ \ \ \ \ \ Interface Quality & 5.16 (1.13) & 5.02 (1.16) & 5.18 (0.64) & 0.106 & 0.005 & .900 &\\
    \ \ \ \ \ \ Possibility to Examine & 4.89 (0.88) & 4.80 (0.81) & 4.47 (0.96) & 0.947 & 0.043 & .396 &\\
    \ \ \ \ \ \ Performance & 4.90 (0.76) & 4.70 (1.11) & 4.57 (1.02)& 0.444 & 0.021 & .644 & \\
    \ \ \ \ \ \ Total & 4.97 (0.52) & 4.98 (0.56) & 4.46 (0.66)& 3.897 & 0.157 & .028 & *\\
    IMI (scale: 0 - 6)\\
    \ \ \ \ \ \ Interest/Enjoyment & 5.37 (0.46) & 5.20 (0.76) & 5.15 (0.53)& 0.557 & 0.026 & .577 &\\
    \ \ \ \ \ \ Pressure/Tension & 3.07 (0.73) & 2.85 (0.59) & 2.27 (0.64)& 5.985 & 0.185 & .005 & **\\
    PXI (scale: 0 - 6)\\
    \ \ \ \ \ \ Challenge & 5.11 (0.85) & 4.98 (0.90) & 3.73 (1.18) & 8.929 & 0.000 & .001 & ** \\
    NASA-TLX (scale: 0 - 99)\\
    \ \ \ \ \ \ Physical Demand & 65.00 (15.81) & 36.00 (23.92) &  25.00 (17.428) & 17.069 & 0.448 & .000 & **\\
    \ \ \ \ \ \ Mental Demand & 53.33 (24.62) & 45.00 (25.00) &  46.33 (21.08) & 0.538 & 0.025 & .588 &\\
   SSQ (scale: 0 - 3)\\
    \ \ \ \ \ \ Nausea & 24.80 (23.31) & 9.54 (13.00) & 41.98 (46.71) & 5.007 & 0.163 & .015 & *\\
    \ \ \ \ \ \ Oculomotor & 20.21 (22.68) & 13.82 (9.92) & 33.37 (29.35)& 3.137 & 0.126 & .062\\
    \ \ \ \ \ \ Disorientation & 22.27 (24.57) & 7.42 (12.74) & 44.54 (46.53)& 5.770 & 0.203 & .009 & *\\
    \bottomrule
     &&&&& \multicolumn{3}{c}{*\textit{p} $<.05$, ** \textit{p} $<.01$}\\
\end{tabularx}
\end{table*}

\begin{itemize}[leftmargin=*]\compresslist
\item H1: In comparison to the Gamepad condition, the Tracker and HMD approaches \textbf{significantly increase the perceived presence}.
\item H2: The full-body conditions require a \textbf{significantly higher physical effort} compared to the Gamepad controls.
\item H3: The Tracker and HMD conditions \textbf{significantly increase task difficulty but not the success rate}.
\item H4: Movement-based sneaking \textbf{significantly boosts players' enjoyment and tension} compared to the Gamepad approach.
\end{itemize}

\noindent Apart from these four hypotheses, we are also interested in how the players perceive the different sneaking techniques: Are the controls easy to learn and usable? How do the techniques compare to real sneaking? Finally, we want to use the insights from the hypotheses, logged gameplay data, and participants' feedback to compare our two proposed approaches against each other. We summarize both aspects into two additional research questions:

\begin{itemize}[leftmargin=*]\compresslist
\item RQ1: How do players assess their particular sneaking technique regarding usability, learnability, and realism?
\item RQ2: Are there notable differences between the two proposed walking-based approaches HMD and Tracker? 
\end{itemize}

\subsection{Scenario}
We realized the stealth game, used for comparing the different sneaking techniques, with the Unity game engine~\cite{unity}. The setting is futuristic and highly technological, including teleports and patrolling robots (see Figure~\ref{fig:scenario}). The players take the role of a spy who must steal a tablet with confidential information hidden somewhere in the level. Since two of our sneaking mechanisms rely on natural walking, we restricted the virtual environment's size to match our real play area's boundaries, i.e., $16 \, m^2$. Each of the ten consecutive levels is structured similarly: The players enter the room using a teleporter, which serves as a loading screen.  While obstacles, e.g., crates, walls, or laser barriers, differ each time, at least two elements are found in every level: the target tablet and a sentinel robot.

This guard is the main antagonist in the game. If players fail to sneak while walking through the level, they attract attention to their position, causing the robot to investigate the noise's source. If the robot detects the players visually, a bar indicating the alertness level begins to fill. Similar to other games, only the HMD's position is used to determine visibility. Players who fail to interrupt the robot's line of sight will be caught and must restart the level. Therefore, players must do their best to stay unseen and unheard. In the case of detection, the environment offers various corners and blinds that help the players hide and wait for the guard to stop searching. Additionally, we provide a holographic noise indicator attached to the hand, informing the players whether they are sneaking quietly enough. As only one condition involves positional foot tracking, we refrain from visualizing feet or body to assure comparability. Also, prior research already covered the effects of displaying virtual limbs~\cite{pan2019foot}.

While walking through the level and locating the tablet, the players have to overcome various challenges and obstacles. These are designed carefully to introduce new gameplay concepts one at a time. In the beginning, players only need to sneak to avoid the guard, who is looking out of a window. Subsequently, we add the guard's lookout, time-based patrols, dynamic obstacles, and crouching activities. A complete list of the mechanisms used in each of the ten levels is depicted in Table~\ref{tab:levels}. After retrieving the tablet, the players must return quietly to their starting point. They are then teleported back to a waiting room where they start the next level.

\subsection{Procedure and Applied Measures}

We conducted a between-subject study splitting the participants randomly into three groups, each using one of the proposed sneaking techniques. In the beginning, we considered a within-subject design to collect qualitative feedback comparing the different implementations. However, the necessary repetition of gameplay elements paired with the general similarity between the sneaking approaches would have led to unwanted sequence effects.

We executed the study in our VR lab using an HTC Vive Pro Wireless setup~\cite{vive}. On average, the study took 45 minutes. At first, the participants were informed about the overall process and completed a general questionnaire assessing gender, age, gaming behavior, and prior VR experience. We also administered the Immersive Tendencies Questionnaire (ITQ)~\cite{Witmer.1998} to determine the ability to get immersed in fiction. At last, we introduced the participants to the VR hardware and assisted them in putting on the hardware.

\begin{table*}[t]
  \caption{Mean scores, standard deviations, and one-way ANOVA values of the custom questions (CQ).}
  \label{tab:custom}
  \begin{tabularx}{\textwidth}
  {>{\raggedright\arraybackslash}p{0.75cm}
  >{\raggedright\arraybackslash}p{5cm}
  >{\centering\arraybackslash}X
  >{\centering\arraybackslash}X
  >{\centering\arraybackslash}X
  >{\centering\arraybackslash}p{1.5cm}
  >{\centering\arraybackslash}p{1.5cm}
  >{\centering\arraybackslash}p{0.6cm}
  >{\raggedright\arraybackslash}p{0.3cm}}
    \toprule
    \multicolumn{2}{l}{Question Item} & Tracker & HMD & Gamepad & F(2,42) & $\eta^{2}$ & $p$ &\\
    \midrule
    CQ1 & The sneaking felt realistic. & 5.20 (1.01) & 4.53 (1.30) & 3.67 (1.50) & 5.361 & 0.203 & .008 & **\\
    CQ2 & The sneaking did not feel right. & 1.80 (2.04) & 1.93 (2.15) & 1.60 (1.55) & 0.113 & 0.005 & .893 &\\
    CQ3 & The sneaking technique was intuitive. & 4.67 (1.50) & 4.67 (1.18) & 5.60 (0.63) & 3.251 & 0.134 & .049 & *\\
    CQ4 & I had to make an effort to sneak. & 4.20 (1.47) & 3.07 (1.62) & 1.00 (0.93) & 27.759 & 0.499 & .000 & **\\
    CQ5 & The sneaking was too difficult. & 2.00 (1.69) & 1.27 (1.16) & 0.20 (0.41) & 12.257 & 0.286 & .000 & **\\
    CQ6 & I felt very active while playing. & 5.33 (0.72) & 5.00 (0.85) & 3.73 (1.91) & 25.886 & 0.238 & .019 & **\\
    CQ7 & I would have preferred another sneaking technique. & 1.80 (1.78) & 0.73 (1.22) & 1.80 (1.52) & 2.445 & 0.104 & .099 &\\
    CQ8 & I would have liked to play more levels. & 5.27 (1.33) & 5.07 (0.88) & 4.53 (2.10) & 0.928 & 0.042 & .403 &\\
    CQ9 & I would like to play more sneaking-based VR games in the future. & 5.13 (1.25) & 4.87 (1.36) & 4.20 (1.42) & 2.026 & 0.088 & .145 &\\
    \bottomrule
     &&&&& \multicolumn{3}{c}{*\textit{p} $<.05$, ** \textit{p} $<.01$}\\
\end{tabularx}
\end{table*}

The participants started the game in a waiting room, where they were introduced to the controls needed to complete the levels. While there was no guard present in the waiting room, the subjects could use their holographic noise indicator to test the particular sneaking technique. We did not limit this introductory phase, which usually only took one to two minutes. After getting used to the controls, the subjects entered the first level by stepping on the teleporting device and returned to the waiting room between every level. We logged the relevant statistics, such as the overall playtime or the number of detections, to analyze the players' performance.

After completing the final level, the subjects removed the HMD and filled out a series of questionnaires regarding their experience. For administering the feeling of presence, we used the Presence Questionnaire (PQ)~\cite{UQO.2004, Witmer.1998} focussing on interaction-related presence. It contains five subdimensions, each rated on a 7-point Likert scale (coded 0 - 6): \textit{realism}, \textit{possibility to act}, \textit{quality of interface}, \textit{possibility to examine}, and \textit{self-evaluation of performance}. We also included the \textit{challenge} construct of the Player Experience Inventory (PXI)~\cite{abeele2020development} and two subscales of the Intrinsic Motivation Inventory (IMI)~\cite{ryan2000self}: \textit{interest/enjoyment} and \textit{pressure/tension} (coded 0 - 6). 

For assessing the mental and physical effort, we used the NASA Task Load Index (NASA-TLX) questionnaire~\cite{hart1988development}. The two chosen subscales \textit{mental demand} and \textit{physical demand} are rated on a 100-point scale with 5-point steps (coded 0-99). Finally, we also included the Simulator Sickness Questionnaire (SSQ)~\cite{Kennedy.1993} with its three subscales \textit{nausea}, \textit{oculumotor}, and \textit{disorientation}. While we expected to find certain differences in cybersickness, these should relate to the different locomotion techniques, i.e., walking and gamepad, and not to the sneaking mechanism itself. The questionnaires were accompanied by a set of custom questions (coded 0 - 6) to gain further insights into the participants' experiences with the sneaking mechanisms (see Table~\ref{tab:custom}). We finished the study by allowing the subjects to share their opinions in a semi-structured interview.

\section{Results}

In total, 45 persons (19 female, 26 male) participated in our study with a mean age of 24.64  (SD=3.13). Most of the subjects played digital games a few times a month ($93\%$) and had already used VR systems before ($78\%$). However, only a minority of $11\%$ reported using VR regularly. All participants were randomly split into three study conditions. We did not find any significant discrepancies between these groups regarding age, gender, prior VR experience, or immersive tendencies (all $p>.05$). As we searched for differences between the three independent groups, we performed one-way analyses of variances (ANOVA) for all measures. Therefore, we ensured normal distribution with Kolmogorov-Smirnov and homogeneity of variances using Levene's tests. In cases where the data did not meet the latter requirement, we used Welch's ANOVA instead. Depending on Levene's test results, we chose either Tukey's or Games-Howell tests for posthoc comparisons. For legibility reasons, we only report significant differences between the conditions, including all necessary information to ensure reproducibility~\cite{vornhagen2020statistical}.

\subsection{Questionnaires}
We assumed that the walking-based implementations lead to higher enjoyment and tension levels. Table~\ref{tab:quest} depicts the resulting scores of the interest/enjoyment and pressure/tension subscales of the IMI. Only the difference in experienced tension is significant, according to the ANOVA ($p=.005$). Posthoc comparisons indicate that the Gamepad condition elicited significantly less tension than the Tracker condition ($p=.005;$ $95\%$ $CI[-1.382,-0.218]$) and the HMD condition ($p=.048;$ $95\%$ $CI[-1.169,-0.005]$).

Moreover, we compared the perceived presence between the three study conditions using the PQ questionnaire. As shown in Table~\ref{tab:quest}, only the measure for experienced realism and the total presence score indicate a significant difference. For perceived realism, the posthoc tests show that the conditions Gamepad and Tracker ($p=.003;$ $95\%$ $CI[-1.626,-0.298]$), as well as Gamepad and HMD ($p=.002;$ $95\%$ $CI[-1.673,-0.346]$) differed significantly. Regarding the total presence, only the difference between Gamepad and HMD ($p=.050;$ $95\%$ $CI[-1.037,-0.001]$) is significant.

Further, we wanted to assure that potential cybersickness induced by locomotion would not impede our study. The results of the SSQ are listed in Table~\ref{tab:quest}. The values for the Gamepad condition are significantly higher than for the HMD condition (nausea: $p=.018;$ $95\%$ $CI[4.882,59.990]$, disorientation: $p=.023;$ $95\%$ $CI[4.999,69.241]$). We presume that this difference is related to the used locomotion technique rather than the sneaking mechanism. Most importantly, the values across all groups are low compared to Kennedy et al.'s reference values~\cite{Kennedy.1993}, thus not indicating significant problems with cybersickness in any of the three conditions.

To test our hypotheses, we also assessed individual subscales of the NASA-TLX and PXI. The resulting means and standard deviations are listed in Table~\ref{tab:quest}. The perceived challenge, according to the PXI, differed significantly across the conditions. The posthoc tests indicate that the Tracker ($p=.001;$ $95\%$ $CI[0.504,2.252]$) and HMD ($p=.004;$ $95\%$ $CI[0.371,2.118]$) conditions provided a greater challenge than the Gamepad controls. For the NASA-TLX, only the physical demand subscale reveals a notable difference: the Tracker condition required a substantially higher physical effort than the two other groups (Gamepad: $p<.001;$ $95\%$ $CI[24.961,55.039]$, HMD: $p=.002;$ $95\%$ $CI[10.525,47.475]$).

\subsection{Custom Questions and Logging Data}
To better understand the participants' expectations and experiences, we also assessed several custom questions. These questions covered three aspects: realism, usability, and liking. Table~\ref{tab:custom} lists the means, standard deviations, and ANOVA results. In particular, the performed one-way ANOVA reveals notable disparities for five of the ten custom questions. For CQ1, posthoc tests indicate that the Tracker mechanism felt more realistic than the Gamepad approach ($p=.006;$ $95\%$ $CI[0.392,2.674]$). Even though the ANOVA suggests a significant difference for CQ3, posthoc testing could not confirm this assumption. 

Concerning CQ4, the Gamepad condition required a substantially lower effort to sneak than the other two groups (Tracker: $p<.001;$ $95\%$ $CI[-4.324,-2.077]$, HMD: $p=.001;$ $95\%$ $CI[-3.278,-0.855]$). Further, the posthoc test reveals a notable difference between the Gamepad condition and the Tracker ($p=.003;$ $95\%$ $CI[-2.962,-0.638]$) and HMD ($p=.010;$ $95\%$ $CI[-1.882,-0.251]$) conditions for CQ5. For CQ6, only the difference between Gamepad and Tracker conditions is significant ($p=.019;$ $95\%$ $CI[-2.945,-0.255]$).

\begin{figure}[tb]
 \centering %
 \includegraphics[width=\columnwidth]{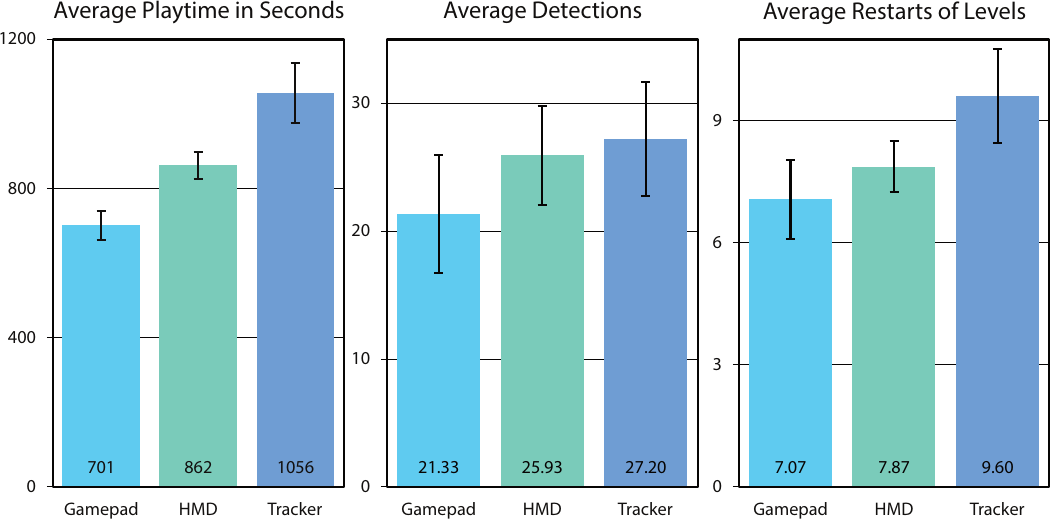}
 \caption{Results from the data logged during the play sessions (means and standard errors). From left to right: The difference in playtime for all study groups in seconds; The average number of audible detections by the NPC guard; The average number of level restarts caused by the players being caught.}
 \label{fig:logs}
\end{figure}

Finally, we also analyzed the data logged during the play sessions: the total playtime, the number of detections by the guard, and the number of level restarts (see Figure~\ref{fig:logs}). Among those, only the playtime differs significantly across the groups. Compared to the Gamepad condition, subjects in the HMD group played $23\%$ longer ($F(2,26.33)=9.410$; $p=.014$; $\eta^{2}
=0.327$; $95\%$ $CI[29.808,293.276]$), and participants using the Tracker approach played even $50\%$ longer ($F(2,26.33)=9.410$; $p=.002$; $\eta^{2}=0.327$; $95\%$ $CI[129.543,580.650]$). However, this discrepancy was not caused by substantial differences in detections or level restarts.

\section{Discussion}

Regardless of the used sneaking technique, the majority of players enjoyed participating in our study. Subjects explicitly mentioned the potential of sneaking-based interactions for immersive experiences. According to their feedback, this type of gameplay conforms to the key advantage of VR by allowing to \textit{"become someone completely different --- like a master spy"(\textbf{P29})}. Most subjects wanted to play more levels (CQ8) and were interested in other sneaking-based VR games (CQ9). But how do the different implementations compare? Where are the individual strengths and weaknesses? Our four hypotheses and two research questions cover the various aspects of player experience and usability necessary to answer these questions.

\medskip
\noindent\textbf{H1: In comparison to the Gamepad condition, the Tracker and HMD approaches significantly increase the perceived presence.}

\noindent The results from the PQ indicate that the subjects experienced high levels of presence across all three conditions. However, the two movement-based techniques both scored higher regarding experienced realism. This finding supports our primary research goal of creating a natural and realistic sneaking experience for IVEs. Interestingly, the HMD condition got even minimal higher values for realism and total presence than the precise tracker-based approach. Players in the HMD group even reported having the feeling that \textit{"the guard would hear every loud step immediately"(\textbf{P5})}. It seems that most participants did not notice the inferior tracking fidelity.

\medskip
\noindent \textbf{H2: The full-body conditions require a significantly higher physical effort compared to the Gamepad controls.}

\noindent One potential difficulty introduced by our proposed mechanisms is the additional physical demand. The requirement of cautiously placing every foot increases the necessary physical effort for the Tracker condition significantly. While the values for the HMD condition were also higher than for the Gamepad group, this difference was not significant. Also, VR experiences do not necessarily suffer from demanding full-body interactions. In contrast, several participants appreciated the novel challenge as it forced them \textit{"to be fully aware of the own body"(\textbf{P34})}. Especially the tracker-based approach caused the subjects to feel very active (CQ6) and was sometimes as demanding as an \textit{"exergame"(\textbf{P1})}.

\medskip
\noindent \textbf{H3: The Tracker and HMD conditions significantly increase task difficulty but not the success rate.}

\noindent The result from the assessed PXI subscale indicates that our two proposed approaches increased the game's challenge significantly. Since both mechanisms are movement-based, players had to stay cautious and canny even in stressful situations. This conflict was especially noticeable when players got detected and had to outsmart the guard to avoid failing the level. Several participants reacted hectically, which sabotaged their goal and usually forced them to restart the level. As a consequence, the players progressed more slowly and carefully. While the subjects' cautious behavior increased the overall playtime, it had no notable effects on the success rate or enjoyment. Since developers typically seek this kind of behavior in stealth games, we consider our approaches' increased challenge to benefit the intended genre. 

\medskip
\noindent\textbf{H4: Movement-based sneaking significantly boosts players’ enjoyment and tension compared to the Gamepad approach.}

\noindent The study's results confirmed our fourth hypothesis only partially. Regarding the participants' enjoyment, we did not find any notable difference between the three groups. Nevertheless, the high scores of the IMI subscale indicate an overall pleasant experience. Especially in the groups using real walking, players sometimes described the gameplay as \textit{"if you were a secret agent in a blockbuster movie"(\textbf{P9})}. Additionally, the tension subscale revealed that our two proposed techniques elicited a significantly higher tension than the gamepad controls. While there is no considerable difference between the more accurate tracker-based mechanism and the more abstracted HMD approach, the results still underline the general advantage of body-based sneaking. These outcomes are also reflected in the oral feedback: \textit{"it is extremely thrilling to play hide-and-seek with the guard after already being detected"(\textbf{P22})}.

\medskip
\noindent\textbf{RQ1: How do players assess their particular sneaking technique regarding usability, learnability, and realism?}

\noindent In general, most participants learned to use the particular sneaking technique very quickly. The concepts behind each mechanism were understood intuitively (CQ3) and did not require additional assistance from the study coordinators. Especially the visual noise indicator, attached to the players' hand, was mentioned as a major aid as it helped to \textit{"gain a feeling for the own loudness"(\textbf{P14})}. After the first levels, the participants mostly \textit{"learned the acceptable threshold by heart"(\textbf{P10})} and reduced their use of the display.

Since all three sneaking techniques work differently, the players' difficulties and problems varied as well. Some of the subjects using the gamepad approach reported being challenged by pressing multiple buttons simultaneously, e.g., taking the tablet while sneaking. In particular, participants with no prior VR experience tended to utter this concern. In the other groups, players had to pay more attention to their movement. Later levels increased this challenge by adding obstacles requiring precise timing or crouching. As this hurdle is primarily relevant for the tracker-based approach, subjects in this group had significantly more problems (CQ5). However, it is worth noting that only very few players reported being overstrained (CQ5) or preferring a different technique (CQ7).

Apart from usability and learnability, we were also interested in the realism of our proposed techniques. How do they compare to real sneaking? In general, players tend to adapt to simplifications and abstractions in gameplay elements and usually ignore implausibilities for the sake of a coherent game experience. Therefore, it does not surprise that the participants mostly rejected CQ2. However, asking them personally, subjects articulated more diversified feedback. Especially the binary sneak mode of the Gamepad condition \textit{"did not feel like sneaking but cheating. I stopped using it and controlled my speed manually"(\textbf{P3})}. For the tracker-based sneaking, players liked that  \textit{"every step is relevant. I could even stomp to attract attention on purpose"(\textbf{P28})}. This feedback is reflected in the significant difference regarding the realism of the sneaking technique (CQ1).

\medskip
\noindent\textbf{RQ2: Are there notable differences between the two proposed walking-based approaches HMD and Tracker?}

\noindent Our proposed implementations fit their intended purpose and provide valuable benefits. Compared to established controls, these approaches can increase the perceived presence, tension, and challenge without causing frustration or exhaustion. However, generally speaking, both proposed techniques performed very similarly, despite the differences in tracking precision. While the accurate tracker-based approach offers some advantages over the HMD abstraction, e.g., more realism and physical activity, these did not significantly impact the player experience. In contrast, the feedback for the Tracker condition was more ambiguous than the overall positive HMD feedback. 

The oral feedback points towards the most likely reason: with the hardware trackers, players must pay attention to every single step they take. In contrast, the HMD approach focuses on the players' intentions. Precisely speaking, the mechanism is based on the assumption that sneaking players walk slowly. Despite being unaware of the actual implementations, players intuitively understood this concept. The HMD approach delivers a comparable experience to precise tracking while requiring less attention and being more forgiving. Therefore, it might fit the players' expectations better.

Furthermore, the HMD implementation relies only on standard hardware and does not require any additional tracking devices. Considering both properties, we conclude that the HMD-based approach is most suited for consumer-oriented experiences, i.e., VR stealth games. In contrast, the hardware trackers can guarantee far more precise footstep tracking necessary for enhanced realism and additional gait-based interactions that are not possible with the HMD abstraction, e.g., stomping. The technique also builds upon Vive's marker-based tracking and is therefore easily applicable to every VR scenario. These benefits make our approach valuable for other use-cases as well. For instance, accurate footstep tracking is useful for a variety of physical activities such as dancing or in specific simulation or training applications.

\section{Limitations}
In our study, we compared our two movement-based approaches against the commonly used gamepad controls. This decision was motivated by the observation that current stealth VR-games tend to rely on button-controlled sneak modes and joystick locomotion. Nevertheless, the chosen study setup raises the question of whether our findings may instead originate from this difference in locomotion techniques. While we indeed attribute a notable influence to the particular navigation method, the overall low levels of cybersickness paired with the similar enjoyment levels indicate a comparable user experience across the groups.

Further, we focussed entirely on the fundamental sneaking mechanism, comprising a simple binary differentiation between walking and sneaking. Consequently, more complex interaction patterns, such as the introduced stomping concept, were not included in the study scenario. Especially stomping is an interesting concept, as it introduces an additional degree of input fidelity, namely exceeding the sneaking-threshold deliberately.  However, such interactions are not easily realizable for the Gamepad and HMD approaches without adding additional buttons. Since this step would have introduced an additional variable for our study, we decided to test the basic sneaking concept in isolation.

Finally, we decided against any avatar visualization except for the players' hands. This decision was mainly motivated by the differences in tracked devices: Only one condition featured positional feet tracking. Therefore, displaying virtual feet would have introduced an additional variable to our study. Instead, we point to the existing work on virtual avatars as the underlying effects were already extensively covered. Similarly, we decided to base our detection algorithms solely on the visibility of the HMD. Thereby, we assured comparability and avoided frustration caused by peeking limbs. Nevertheless, using the players' fully tracked and visualized bodies for the detection mechanism might significantly influence the players' behavior and the measured effects.

\section{Conclusion and Future Work}

Virtual reality games provide the unique opportunity to slip into the role of the own favorite character and experience a thrilling adventure first-handed. One of these stories may involve a top-secret agent on his almost impossible mission to save the world. Therefore, the agent must infiltrate a restricted area and secure confidential files. While players can already encounter similar plots in available VR games, these often lack to convey a satisfying stealth experience. In most cases, players only need to stay out of sight of the probably deaf guards. Stepsounds rarely have any influence on the gameplay. This simplification does not use the full interaction fidelity of sneaking activities and limits the achievable scope of realism.

With our work, we have explored the potential of sneaking as a novel input modality for such IVEs. Therefore, we developed two gait-oriented mechanisms to capture the users' gait. The first measured the feet's deceleration, while the second used the average HMD speed as a proxy. We then compared the two approaches against the established gamepad controls. Our study's results revealed three interesting takeaways:
\begin{enumerate}
\item The experiments confirmed that players generally appreciated sneaking-based gameplay elements.
\item Our proposed implementations increased the perceived presence, tension, challenge, and physical activities without overcharging or exhausting the players. 
\item Both approaches performed very similarly, despite the differences in tracking fidelity and degree of abstraction. In most cases, it is not necessary to capture exact foot movements, as the HMD-based implementation provides a similar experience.
\end{enumerate}

\noindent In our future research, we will concentrate on further use-cases of the proposed interaction concepts. In particular, our tracker-based mechanism opens interesting research directions. It enables precise detection of the users' steps, which could be used for other movement-intense activities, such as dancing. While we mainly focused on differentiating walking and sneaking, our tracking mechanism was also capable of detecting stomping and jumping movements. In the future, we want to extend this concept to identify more types of gait. Especially training simulations could profit from this capability to provide individual gait-understanding feedback. Finally, we aim to combine the proposed gait-based sneaking with additional channels, namely synchronized audio feedback, voice detection, haptic feedback, and full-body tracking, to achieve a fully lifelike sneaking experience.

\bibliographystyle{abbrv-doi-hyperref-narrow}
\balance

\bibliography{template}

\end{document}